\documentclass[useAMS,usenatbib,fleqn]{mn2e}
\usepackage[usenames,dvipsnames]{color}
\usepackage{times}
\usepackage{mathptmx}
\usepackage{graphicx}
\usepackage{gensymb}
\usepackage[table]{xcolor}
\usepackage{etoolbox}
\usepackage{xspace}
\usepackage{paralist}
\usepackage[usenames,dvipsnames]{color}
\usepackage{amsmath, bm}

\begin{document}
\newtoggle{comments}
\newtoggle{changed}
\newtoggle{guideline}

\toggletrue{comments}
\toggletrue{changed}
\toggletrue{guideline}

\newcommand{\comment}[1]{{\iftoggle{comments}{ \color{BrickRed} #1}{}}}
\newcommand{\changed}[1]{{\iftoggle{changed}{ \color{NavyBlue}}{} #1}}
\newcommand{\guideline}[1]{{\iftoggle{guideline}{ \color{PineGreen}}{} #1}}

\newcommand{\kpc}{\,\rm{kpc}\xspace}
\newcommand{\kms}{\,\rm{km\,s^{-1}}\xspace}
\mathchardef\mhyphen="2D
\newcommand{\nmagic}{\textsc{nmagic}\xspace}
\newcommand{\brava}{\textsc{brava}\xspace}
\newcommand{\vvv}{\textsc{vvv}\xspace}
\newcommand{\Msun}{\,\rm{M}_{\odot}\xspace}
\newcommand{\R}{\mathcal{R}}
\newcommand{\Gyr}{\,\rm{Gyr}}
\newcommand{\reftable}{\rm{Table}\xspace}
\newcommand{\reffigure}{\rm{Fig.}\xspace}
\newcommand{\refequation}{\rm{Eq.}\xspace}
\newcommand{\refequations}{\rm{Eqs.}\xspace}
\newcommand{\refsection}{\rm{Section}\xspace}

\title[Peanuts, brezels and bananas]{Peanuts, brezels and bananas:\\ food for thought on the orbital structure of the Galactic bulge}
\author[M. Portail et al.]
  {M.~Portail,$^1$\thanks{E-mail:portail@mpe.mpg.de}
  C.~Wegg$^1$ and O.~Gerhard,$^1$\\
  $^1$ Max-Planck-Institut f\"{u}r Extraterrestrische Physik, Gie\ss enbachstra\ss e, D-85741 Garching, Germany}

\date{Accepted 2015 March 24.  Received 2015 March 19; in original form 2015 February 19}
\pagerange{\pageref{firstpage}--\pageref{lastpage}} \pubyear{2015}
\maketitle
\label{firstpage}

\begin{abstract}
Recent observations have discovered the presence of a Box/Peanut or X-shape structure in the Galactic bulge. Such Box/Peanut structures are common in external disc galaxies, and are well-known in N-body simulations where they form following the buckling instability of a bar. From studies of analytical potentials and N-body models it has been claimed in the past that Box/Peanut bulges are supported by ``bananas'', or $x_1\rm{v}_1$ orbits. We present here a set of N-body models where instead the peanut bulge is mainly supported by brezel-like orbits, allowing strong peanuts to form with short extent relative to the bar length. This shows that stars in the X-shape do not necessarily stream along banana orbits which follow the arms of the X-shape. The brezel orbits are also found to be the main orbital component supporting the peanut shape in our recent Made-to-Measure dynamical models of the Galactic bulge. We also show that in these models the fraction of stellar orbits that contribute to the X-structure account for $40-45\%$ of the stellar mass.
\end{abstract}


\section{Introduction}
\label{section:Introduction}
There has been much recent interest in the dynamical structure of the Galactic bulge, following the discovery of the bimodal distribution of red clump giants (RCGs) magnitudes in upper bulge fields \citep[the split red clump;][]{McWilliam2010,Nataf2010,Ness2012}. Analyzing 50 million RCGs from the \vvv survey, \citet{Wegg2013} were able to measure directly the 3D density distribution of the dominant bulge stellar population, finding a very pronounced Box/Peanut bulge (B/P bulge) structure extending about $2\kpc$ along its major axis. B/P bulges are found in about half of the external edge-on disc galaxies \citep{Lutticke2000} and are the focus of several recent and on-going studies \citep{Williams2011, Fabricius2012,Walcher2014, Seidel2015a}. B/P bulges are common in N-body simulations of barred discs where instability of the bar leads to one or more buckling events, creating a prominent peanut shape in the inner part of the bar \citep{Combes1981, Raha1991,Martinez-Valpuesta2006}.

The orbital structure of bars was first studied in two dimensions \citep[for a review see][]{Contopoulos1989}, showing that bars are largely made out of regular orbit families trapped around stable periodic orbits. Later, studies of the 3D orbital structure were carried out, focusing mostly on the stable periodic orbit families. The third dimension introduces instability in the dynamics and leads to bifurcations of stable periodic planar orbits. Pioneering work by \citet{Pfenniger1991} found that the main orbit family of planar bars (the $x_1$ orbits) becomes vertically unstable in certain regions and bifurcates to several 3D orbital families including the $x_1\rm{v}_1$ family. $x_1\rm{v}_1$ orbits are also called banana orbits because of their banana shape when seen side-on. These banana orbits have since been considered as the backbone of B/P bulges \citep[eg.][]{Pfenniger1991, Martinez-Valpuesta2006}, even though other higher order resonant orbit families could also build a peanut shape bulge \citep{Patsis2002}.

In this letter we show that the banana orbits are not necessarily the main building block of B/P bulges, challenging the picture of stars streaming along the arms of the X-shape. We describe a new family of brezel-like orbits with which a peanut bulge can form at smaller radii relative to the bar than is possible with banana orbits. We find that this orbit family dominates the peanut shape of the Galactic bulge in both our N-body models (\refsection \ref{section:orbitsNbodyModels}) and our Made-to-Measure models (\refsection \ref{section:orbitsM2MModels}) for the Galactic bulge from \citet{Portail2015}. We also present an orbit-based characterization of the X-shape structure of the Galactic bulge and show that in our models, the Galactic X-shape structure accounts for $40-45\%$ of the stellar mass of the bulge.

\section{The orbital structure of B/P bulges in N-body models}
\label{section:orbitsNbodyModels}
\subsection{N-body models of B/P bulges}
\label{section:initialModels}
We analyze N-body models of barred discs evolved from a near equilibrium stellar disc embedded in different live dark matter halos. During this evolution the disc naturally forms a bar which rapidly buckles out of the Galactic plane and creates a B/P bulge \citep{Combes1981, Raha1991}. In this work we use the initial N-body models M80, M85 and M90 already presented in \citet{Portail2015}. These three models differ by their dark matter fraction in the bulge, ranging from $30\%$ for model M80 to $12\%$ for model M90. Using the definition of \citet{Debattista2000} where $\R$ is the ratio between corotation radius and half-length of the bar, they  span the complete range of reasonable values for the $\R$ parameter \citep{Elmegreen1996a}, from a slow bar for M80 ($\R = 1.8$) to a fast bar for M90 ($\R = 1.08$).

We assume a distance to the Galactic centre of $R_{0} = 8.3\kpc$ \citep{Reid2014, Chatzopoulos2014} and place the bar at an angle of $27\degree$ with respect to the line-of-sight towards the Galactic center \citep{Wegg2013}. We scale the models to the Milky Way by placing the end of their long bar component at $l=27\degree$ as seen from the Sun's location \citep{Hammersley2000, Cabrera-Lavers2007}, resulting in a bar half length of $4.6\kpc$. In the following we use the $(x,y,z)$ coordinates in the rotating frame of the bar where $x$ is along the major axis of the bar and $z$ is the axis orthogonal to the galactic plane. 

\subsection{Orbit classification}
\label{section:orbitClassification}
Most studies of the orbital structure of B/P bulges focus on studying families of periodic orbits in rotating potentials, either analytical \citep[e.g. ][]{Skokos2002} or N-body \citep[][but see also \citet{Harsoula2009}]{Pfenniger1991}. In this paper, we study the orbits of the particles in the N-body potential frozen well after the formation and buckling of the bar, which is then assumed to rotate at a constant pattern speed. We integrate the orbits using a drift-kick-drift adaptive leap-frog algorithm and use frequency analysis to classify those that build the B/P bulge. We consider all stellar particles in a box of $\pm 4\kpc \times \pm 1.5\kpc \times \pm 1.5\kpc$ in the bar frame ($\sim 7 \times 10^5$ particles) and integrate the orbits for $50$ dynamical times, where the dynamical time is defined as the time necessary to complete a circular orbit at $2\kpc$. We record $100$ positions of the selected particles per dynamical time and construct this way a time series for the particle coordinates. For each particle we compute the Fast Fourier Transform of its coordinate time series and identify the main frequency of each coordinate as the frequency corresponding to the highest spectral peak. We do this for the Cartesian coordinates $x$ and $z$ in the bar frame as well as the cylindrical radius $r$ and consider the ratios of the main frequencies $f_r/f_x$ and $f_z/f_x$.

In our models we find two main groups of particles: bar particles and disc particles. Bar particles are identified by their frequency ratio $f_r/f_x \in 2 \pm 0.1$. They have an elongated shape along the bar major or intermediate axis. Disc particles are defined through $f_r/f_x \not\in 2 \pm 0.1$. They do not support the bar shape and do not show any prominent peanut shape. As we focus on the peanut shape of the bar, we exclude the disc particles from further analysis.

\begin{figure}
\centering
\includegraphics{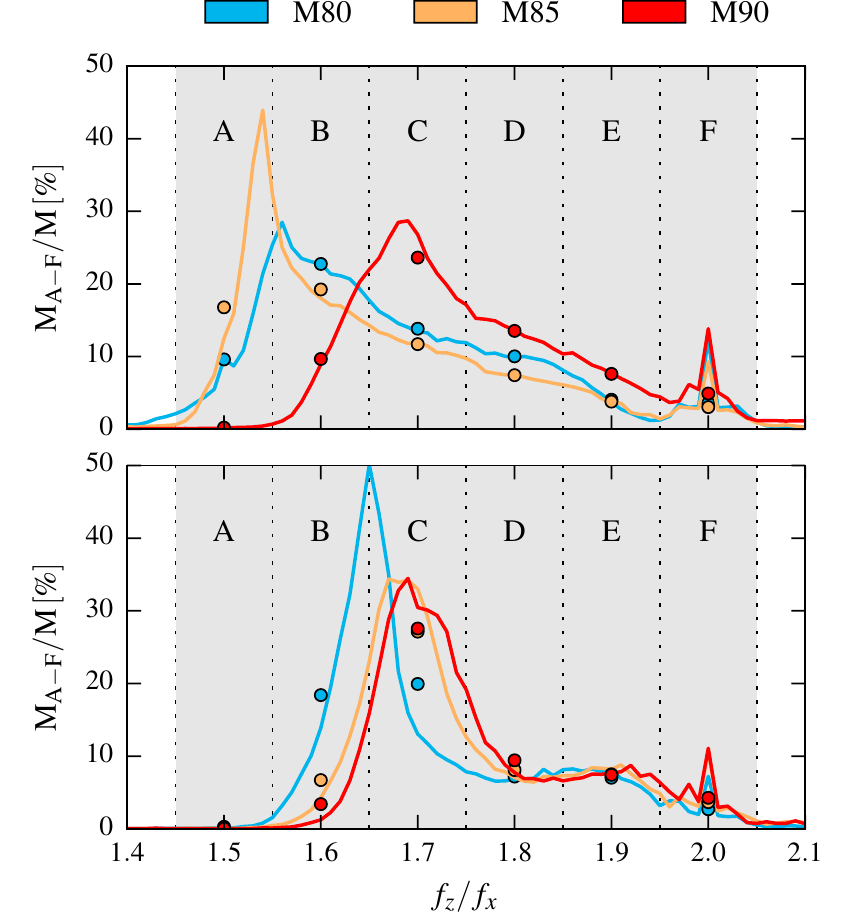}
\caption{Fraction of the selected stellar particles belonging to orbital classes A-F defined through their frequency ratio $f_z/f_x$ as shown in the figure. Top: for the three N-body models (see \refsection \ref{section:orbitsNbodyModels}) ; bottom: for the models fitted to the Milky Way bulge (see \refsection \ref{section:orbitsM2MModels}). The colored dots indicate the fraction of mass in the orbital classes while the solid lines show the continuous distribution of the frequency ratio $f_z/f_x$. The banana orbits are located in the peak at$f_z/f_x = 2$.}
\label{fig:histograms}
\end{figure}

In order to identify the orbital components of the peanut structure we focus on the frequency ratio of the side-on coordinates, $f_z/f_x$. The top plot of \reffigure \ref{fig:histograms} shows the distributions of the frequency ratio $f_z/f_x$ for our three N-body models. As $f_z/f_x$ is almost entirely contained between $1.5$ and $2.0$, we make a low resolution orbit classification by splitting the particles into $6$ classes A to F, corresponding to equally spaced bins in $f_z/f_x$ centreed on $1.5$, $1.6$, $1.7$, $1.8$, $1.9$ and $2.0$. The colored dots in \reffigure \ref{fig:histograms} indicate the fraction of stellar mass belonging to each orbital class in the three models.

\begin{figure}
\centering
\includegraphics{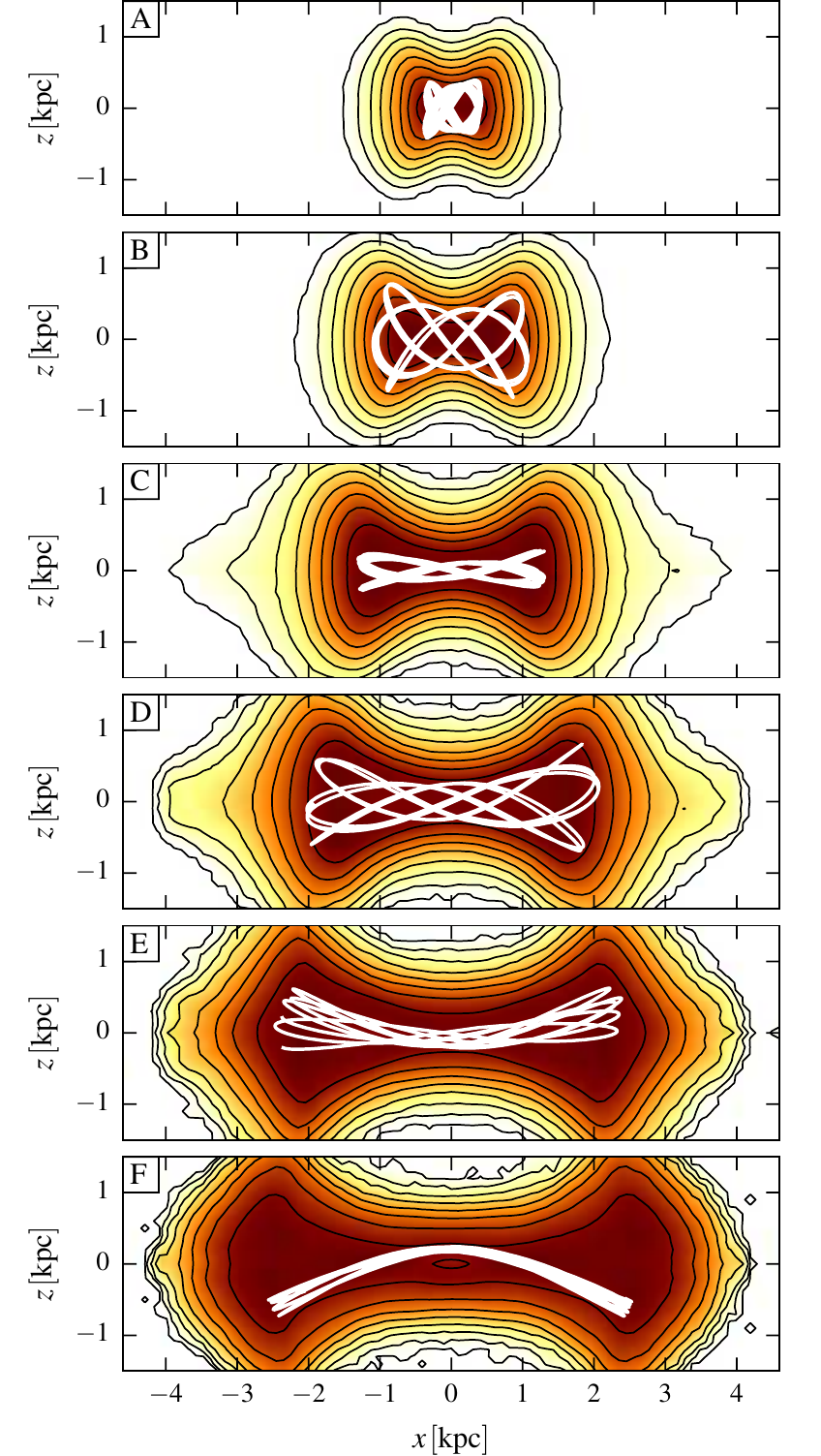}
\caption{Side-on projections of the six orbital classes identified by frequency analysis of model M85. The white curves show the side-on trajectories of typical sample orbits belonging to each class. The banana orbits identified by \citet{Pfenniger1991} are in class F.}
\label{fig:classes}
\end{figure}

\reffigure \ref{fig:classes} shows the side-on projections of these $6$ orbital classes identified in model M85. All these classes display a strong peanut shape on the side-on projection. We find that orbits with larger frequency ratio $f_z/f_x$ have larger radial extent. The total B/P structure of the model is then the sum of all these embedded peanut shapes. All orbits in any one class are very similar to each other and the white curves in \reffigure \ref{fig:classes} show the side-on trajectory of a typical sample orbit of each class. 

\subsection{Which orbit classes dominate the X-shape?}
\label{section:orbitXShape}

The orbital structure of 3D bars has been studied both in analytical potentials by \citet{Contopoulos1994} and \citet{Skokos2002}, and in self gravitating N-body models by \citet{Combes1990} and \citet{Pfenniger1991}. These authors found that the dominant $x_1$ orbit family of 2D bars may become vertically unstable when introducing the third dimension. Instability of planar $x_1$ orbits results in the birth of many different 3D periodic orbits families, called the $x_1$ tree. The simplest member of the $x_1$ tree is the $x_1v_1$ family, also called ``banana'' orbits by \citet{Pfenniger1991} because of the shape of their side-on trajectories. Banana orbits have been found in many N-body or analytical models and have since been considered as the main building block of B/P bulge \citep{Combes1990, Pfenniger1991, Patsis2002a, Martinez-Valpuesta2006, Quillen2013}. They appear at the vertical $2:1$ resonance (i.e. $2$ vertical oscillations per revolution), causing the peak at $f_z/f_x = 2$ in {\reffigure \ref{fig:histograms}}, and are therefore the main contribution to our orbital class F. Higher order members of the $x_1$ tree can also play a role in 3D bars and eventually contribute to the peanut shape, as noted by \citet{Skokos2002} and \citet{Patsis2002a}. However these orbits arise at larger energies, which means even larger radial extent.

Identifying the building blocks of the B/P feature from our N-body models is not a trivial task. B/P bulges are composite structures where different orbits existing at different energies and radii overlap and all together create the B/P feature. In our models, all orbital classes shown in \reffigure \ref{fig:classes} are at some radius the main component of the 3D part of the bar. In order to clarify this, we make the distinction between Box/Peanut (B/P) shape and X-shape. B/P shape refers to the shape of the isophotes in the side-on projection, while X-shape refers to the typical X-feature that is revealed when applying unsharp-masking techniques to the side-on projection of the B/P feature. Determining the X-shape is a common practice to study external galaxies and can be performed in several ways, using a median filter as in \citet{Bureau2006} or removing a fit of the image, as in \citet{Li2012}.

\begin{figure}
\centering
\includegraphics{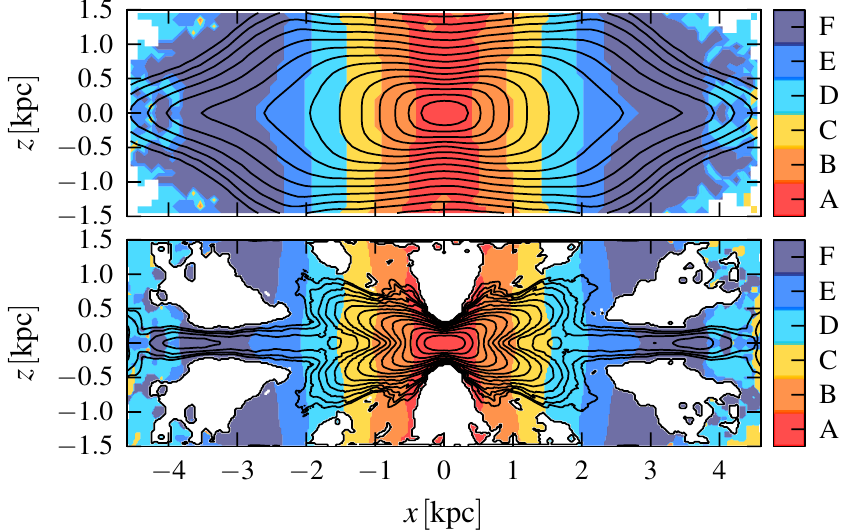}
\caption{Peanut shape (top) and X-shape produced by unsharp-masking (bottom), for model M85. The colors indicate the orbital class which contribute the highest surface density at each point of the side-on view.}
\label{fig:coloredXshape}
\end{figure}

\reffigure \ref{fig:coloredXshape} shows in black contours the B/P shape of the N-body model M85 and its X-shape obtained after removing the median filtered image. In this figure, the colors indicate which orbital class contributes the highest surface density at each point of the side-on view of the bar. The B/P feature, which is the sum of the different orbital classes shown in \reffigure \ref{fig:classes}, appears mostly boxy with no well defined length. On the contrary the X-shape is clear and contained inside $2\kpc$. Only classes A, B, C and D contribute significantly to the X-shape in this model. The banana orbits, the lowest energy member of the $x_1$ tree, dominate the light only at radii larger than $2.5\kpc$ and contain only a small fraction of the stellar mass in the bulge as shown in {\reffigure \ref{fig:histograms}}. Therefore, in this N-body model, the picture that the X-shape is the result of stars trapped around banana orbits streaming along the arms of an X is incorrect. The X-shape in this model consist mostly of similar non resonant orbits whose parent orbits we call \emph{brezel} orbits, which build an X-shape at shorter radii. Brezel orbits are described in \refsection \ref{section:brezel} in the context of the Galactic bulge.

\section{The orbital structure of the Galactic B/P bulge}
\label{section:orbitsM2MModels}
\subsection{Made-to-Measure models of the Galactic bulge}
\label{section:M2MGB}
In \citet{Portail2015} we constructed a set of dynamical models of the Galactic bulge, with different dark matter halos. We used the Made-to-Measure method \citep{Syer1996, DeLorenzi2007} to adapt the particle weights of the N-body models M80, M85 and M90 just described, to force them to reproduce the stellar density and kinematic measurements of the Galactic bulge. We constrained the stellar density using the 3D density map of RCGs in the bulge from \citet{Wegg2013}, which trace the dominant fraction of the stellar mass. This map was derived by deconvolution of extinction and completeness corrected line-of-sight magnitude distributions from the \vvv survey \citep{Saito2012} and covers a box of $\pm 2.2\kpc \times \pm 1.4\kpc \times \pm 1.2\kpc$. As kinematic constraints we used the radial velocity and dispersion measurements from the \brava survey \citep{Kunder2012}. After the Made-to-Measure fit, the resulting three models match the data very well and are equally good candidates to represent the dynamics of the Galactic bulge, differing from each other by their dark matter fraction in the bulge.

\begin{figure}
\centering
\includegraphics{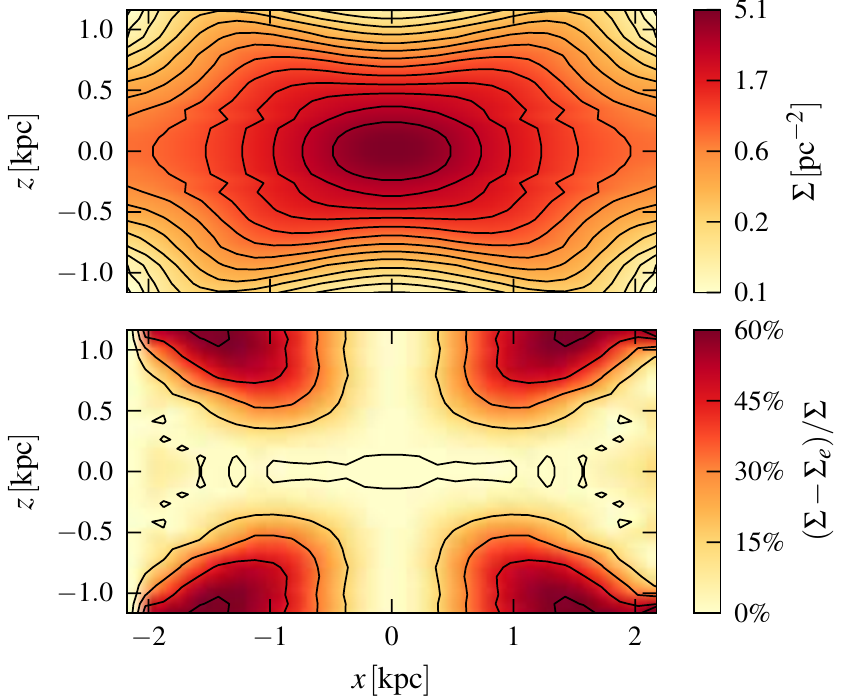}
\caption{\emph{Top:} Side-on projection of the extrapolated 3D red clump giants number density map originally from \citet{Wegg2013}. \emph{Bottom:} Non-elliptical residual fraction of the side-on projection.}
\label{fig:Xshape}
\end{figure}

The top plot of \reffigure \ref{fig:Xshape} shows the side-on projection of the RCGs density map of \citet{Wegg2013} extrapolated in the midplane by \citet{Portail2015}. The peanut shape is strong and mostly contained inside $1.5\kpc$ from the centre along the major axis of the bar. 
It can be highlighted in a more physical way than usually done with unsharp masking. For this we follow \citet{Portail2015} and compute the elliptical component $\Sigma_e$ of the side-on projection of the map $\Sigma$, by fitting a family of ellipses to $\Sigma$ with the constraint that the residuals $\Sigma - \Sigma_e$ are everywhere positive. The non-elliptical residual fraction $(\Sigma - \Sigma_e)/\Sigma$ is plotted on the bottom panel of \reffigure \ref{fig:Xshape}. It shows a large deviation from elliptical shape at radii as short as $1.5\kpc$, about one third of the bar length. In these models, the banana orbits do not achieve significant height at radii around one third of the bar length. The lower plot of \reffigure \ref{fig:histograms} shows the orbital compositions of our three models after fitting to the Milky Way data. The comparison with the upper plot shows that the Made-to-Measure procedure (i.e. the data) moved most of the mass to the orbital classes B and especially C. These classes display a strong peanut at radii around a third of the bar length, as shown in \reffigure \ref{fig:classes}.

Thus, the relation between the X-shape and the orbital structure is not straightforward. X-shapes in external galaxies are often off-centered, with the two arms of the X not crossing at the center as shown by \citet{Bureau2006}. These authors suggested that off-centered X-shapes could be associated with banana orbits given their morphological similarity. The X-shapes of our Made-to-Measure models of the MW bulge are also off-centered (as shown in Fig. 17 of \citealt{Portail2015}), even though the banana orbits represent only a small fraction of the orbital structure in these models.

\subsection{The brezel orbit}
\label{section:brezel}

\begin{figure}
\centering
\includegraphics{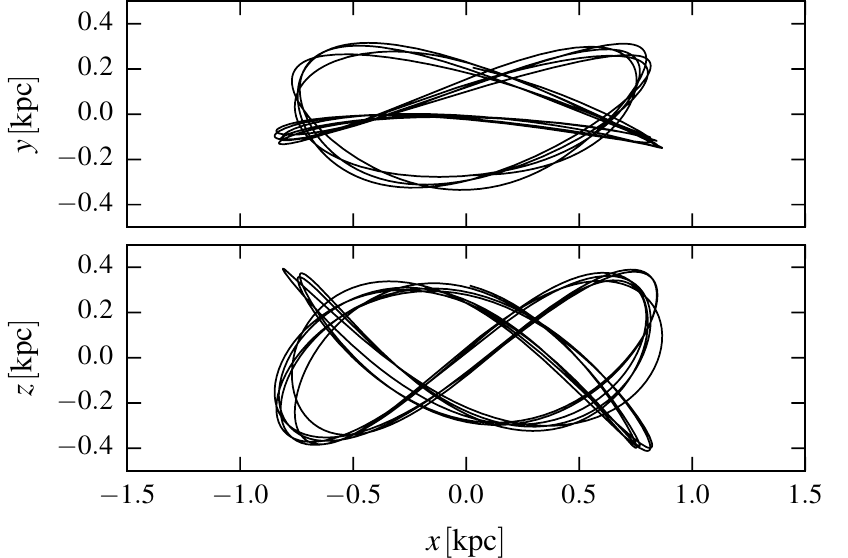}
\caption{Brezel orbit viewed face-on (top) and side-on (bottom).}
\label{fig:brezels}
\end{figure}

After fitting of the \brava data and RCGs density, most of the stellar mass of bar particles are on orbits of class C. These orbits were already present in the initial models but in lower proportion. This class is mostly made of non-resonant orbits which show morphological similarities with one particular orbit family. \reffigure \ref{fig:brezels} shows the face-on (top) and side-on (bottom) view of such an orbit that we call brezel orbit.\footnote{Note that our brezel orbit is topologically different from the ``prezel boxlet'' orbit, found at the 4:3 resonance in triaxial non-rotating potentials \citep{Miralda-Escude1989, Merritt1996}.}
This orbit has a mirror-symmetric counterpart with respect to the $y=0$ plane. It has $f_z/f_x \simeq 5/3$ but needs $10$ oscillations in $z$ and $6$ oscillations in $x$ in order to close. A possible origin for the brezel orbit would be the $5:3$ vertical resonance of the so called $x_1\rm{mul}_2$ planar orbit, in the notation of \citet{Skokos2002} (i.e. the usual 2D $x_1$ orbit but considered to close after two rotations instead of one). Their face-on shapes are also similar to the rm21 and rm22 identified by \citet{Patsis2014a} (see their Fig. 2), arising from a radial bifurcation of the $x_1\rm{mul}_2$. A detailed study of the $x_1\rm{mul}_2$ tree is needed to study this further.

\subsection{The Galactic X-structure}

What is the stellar mass fraction associated with the X-shape in the Galactic bulge? In models where there is no correspondence between the X-shape and a particular orbit family, this question cannot be answered easily. 
In these models, the X-shape only reveals the excess of light over some smooth background in the side-on projection, but is no longer directly related to any physical substructure of the bulge. 
Therefore, using our orbit classification, we define the \emph{X-structure} as the physical stellar component which is built by all bar orbits ($f_r/f_x \in 2 \pm 0.1$) that contribute significantly to the non-elliptical fraction identified in \reffigure \ref{fig:Xshape}; specifically, by all bar orbits that visit regions of the side-on view where the non-elliptical fraction is larger than $10\%$. With this definition, the peanut shape of \reffigure \ref{fig:Xshape} can be seen as the sum of three components: the surrounding disc, made of orbits that do not follow the bar; the main bar, made of orbits that follow the bar but do not contribute significantly to $\Sigma - \Sigma_e$; and the X-structure, contributing to most of the deviation of $\Sigma$ from elliptical density.

With this definition, the fraction of the stellar mass in the X-structure contained in a box of $\pm 4\kpc \times \pm 1.5\kpc \times \pm 1.5\kpc$, is $44\%$, $43\%$ and $42\%$, respectively, for the fitted models M80, M85 and M90. Restricting this computation to the box of $\pm 2.2\kpc \times \pm 1.4\kpc \times \pm 1.2\kpc$ where \citet{Wegg2013} measured the RCGs density, gives very similar numbers, $45\%$, $44\%$ and $43\%$, respectively. This large fraction is in agreement with the lower bound estimated in \citet{Portail2015}. It is mostly due to the morphology of class B and C orbits, which provide the peanut shape as well as a significant part of the in-plane density.

\section{Conclusion}
We have analyzed the orbits of three B/P bulges in N-body models of bars evolved in different dark matter halos. We find that the B/P bulges of these models are complex structures made of the superposition of several peanut shapes produced by different orbits, that are embedded at different radii. Contrary to what is usually stated in the literature, these B/P bulges are not mainly made out of ``banana'' or $x_1\rm{v}_1$ orbits and their X-shape is not the result of stars streaming along these orbits as they follow the arms of the X-shape. Instead, in these models a strong peanut shape bulge is built from a family of brezel-like orbits, with a typical extent of about one third of the bar length.

In our Made-to-Measure models fitted to Milky Way observables, i.e., to the red clump giants 3D density map from \citet{Wegg2013} and the \brava kinematics data \citep{Kunder2012}, we find that the brezel orbit family is the backbone of the orbits populating the Galactic Box/Peanut feature. Using our orbital classification we propose a definition of the physical structure associated with the Galactic X-shape and we estimate that the fraction of stellar orbits that contribute to it account for $40-45\%$ of the stellar mass of the bulge.

\section*{Acknowledgement}
We thank Panos Patsis for helpful discussions on orbits, stability and chaos as well as Mat\'ias Bla\~na for his help in setting up our N-body models of barred discs.


\bsp
\label{lastpage}
\end{document}